Superconductivity in oxygen-annealed FeTe$_{1-x}$S$_x$ single crystal

Yoshikazu Mizuguchi[1,2,3], Keita Deguchi[1,2,3], Yasuna Kawasaki[1,2,3], Toshinori Ozaki[1,2], Masanori Nagao[4], Shunsuke Tsuda[1,2], Takahide Yamaguchi[1,2] and Yoshihiko Takano[1,2,3]

1. National Institute for Materials Science, 1-2-1 Sengen, Tsukuba, 305-0047, Japan
2. Japan Science and Technology Agency-Transformative Research-Project on Iron-Pnictides (JST-TRIP), 1-2-1 Sengen, Tsukuba, 305-0047, Japan
3. University of Tsukuba, 1-1-1 Tennodai, Tsukuba, 305-8571, Japan
4. University of Yamanashi, 7-32 Miyamae, Kofu, 400-8511, Japan

Abstract
We investigated the S-doping-driven phase transition from antiferromagnetic to superconducting in FeTe$_{1-x}$S$_x$ single crystals. The partial substitution of Te by S suppresses antiferromagnetism in Fe-square lattice. Superconductivity is induced by oxygen annealing for only FeTe$_{1-x}$S$_x$ in which the long-range magnetic ordering is suppressed. To realize superconductivity in FeTe$_{1-x}$S$_x$, both S concentration enough to suppress antiferromagnetism and oxygen annealing are required. Anisotropy of superconductivity in oxygen-annealed FeTe$_{0.886}$S$_{0.114}$ was estimated to be 1.17.

I. Introduction

FeTe$_{1-x}$S$_x$ is one of the Fe-chalcogenide superconductors with transition temperature $T_c \sim 10$ K [1-3]. Although the parent phase FeTe exhibits antiferromagnetic ordering below 70 K, a partial substitution of Te by S suppresses the magnetic ordering and induces superconductivity. The FeTe$_{1-x}$S$_x$ superconductor is one of the potential compounds for application among Fe-based superconductors due to its simple composition and comparably low toxicity among the Fe-based superconductors. However, the synthesis of FeTe$_{1-x}$S$_x$ that shows bulk superconductivity is difficult due to the low solubility limit of S for the Te site caused by a large difference in the ionic radius between S and Te [4].

Polycrystalline FeTe$_{0.8}$S$_{0.2}$ does not show bulk superconductivity just after synthesis using the solid-state reaction method. However, superconductivity is induced by exposing the sample to the air. Heating the sample in water, ethanol and alcoholic drinks also induce superconductivity [5,6]. Furthermore, it was found that the better way to quickly induce superconductivity in FeTe$_{0.8}$S$_{0.2}$ is oxygen annealing at 200 °C [7]. To clarify a cause of the evolution of superconductivity by those post treatments, systematic studies using high-quality single crystals are required. We investigated the oxygen annealing effect on physical and structural properties of FeTe$_{1-x}$S$_x$ single crystals with several S concentrations. Here we discuss the roles of S and oxygen for inducing superconductivity in FeTe$_{1-x}$S$_x$.

II. Experimental Details

Single crystals of FeTe$_{1-x}$S$_x$ were synthesized using a self-flux method. Fe powder (99.9 %), S grains (99% up) and Te grains (99.999 %) with several nominal compositions ($x$ = 0.05, 0.1, 0.15, 0.2, 0.25, 0.3, 0.4 and 0.5) were placed in an alumina crucible and sealed into an evacuated quartz tube. The samples were heated at 1050 °C for 20 hours and cooled down to 650 °C with a rate of –4 °C/hour. The obtained single crystals were sealed into a quartz tube filled with atmospheric pressure of oxygen gas, and heated at 200 °C for 2 hours. The actual composition of the crystal was determined using electron probe micro analyzer (EPMA). The value was estimated using an average of four measurements. Detailed information of crystal characterization is summarized in Ref. 4. The lattice constant $c$ for the obtained crystals was estimated by x-ray diffraction using the $2\theta$-$\theta$ method with Cu-K$\alpha$ radiation. Temperature dependence of in-plane resistivity from 300 K to 2 K was measured using a four-terminal method. To discuss anisotropy of superconductivity, temperature dependence of resistivity was measured under magnetic fields up to 7 T both parallel and perpendicular to the $c$ axis of the crystal. Temperature dependence of magnetic susceptibility from 15 to 2 K was measured using a superconducting quantum interference device (SQUID) magnetometer with an applied field of 10 Oe.

III. Results and discussion

Figure 1 is an optical-microscope image of typical oxygen-annealed $FeTe_{0.886}S_{0.114}$ single crystals. The surface of the oxygen-annealed crystal is still shining as well as before annealing. The four terminals for the resistivity measurement were attached on the as-annealed surface.

Figure 2(a) and (b) show the temperature dependence of resistivity for as-grown and oxygen-annealed $FeTe_{1-x}S_x$, respectively. The anomaly corresponding to the magnetic transition, observed in FeTe around 70 K, is suppressed with increasing S concentration. The magnetic transition temperatures for oxygen-annealed $FeTe_{0.97}S_{0.03}$ and $FeTe_{0.952}S_{0.048}$ are almost the same as those before annealing, respectively, which indicates that oxygen annealing does not affect the magnetic transition temperature. In fact, these samples did not show zero resistivity even after oxygen annealing, suggesting that oxygen annealing cannot suppress long-range ordering.

Although the magnetic transition seems to be suppressed by the S substitution for $x$ = 0.064 and 0.114, zero resistivity is not observed for as-grown crystals as shown in Fig. 2(a). By annealing those as-grown crystals in oxygen, superconducting transition is induced as shown in Fig. 2(b). Figure 3 is an enlargement of temperature dependence of resistivity below 30 K. Zero resistivity temperature is estimated to be $T_c^{zero}$ = 7.0 K for both $x$ = 0.064 and 0.114. Interestingly, the superconducting transition temperature for $x$ = 0.064 and 0.114 are almost the same, in spite of the difference in S concentration. This implies that $T_c$ in this system is essentially independent on the S concentration. The $T_c$ of Fe-based superconductor is much sensitive to the local crystal structure; the anion height from the Fe-square lattice is one of the key factors that determine the $T_c$ [8-10]. Furthermore, the valence of S and Te is the same value of –2; S substitution for Te does not change the total charier density. Therefore, $T_c$ of $FeTe_{1-x}S_x$ would not be sensitive to S concentration. Figure 4 shows the temperature dependence of magnetic susceptibility for the oxygen-annealed $FeTe_{1-x}S_x$ crystals. Diamagnetic signal corresponding to superconducting transition is observed below 7.5 K only for $x$ = 0.064 and 0.114. The $T_c^{mag}$ estimated from magnetic susceptibility is almost the same as $T_c^{zero}$ in Fig. 3.

Figure 5(a) and 5(b) shows the temperature dependence of in-plane resistivity for oxygen-annealed $FeTe_{0.886}S_{0.114}$ under the magnetic fields up to 7 T perpendicular and parallel to $c$ axis, respectively. The suppression of superconductivity by magnetic field of $H//c$ is more remarkable than that under $H//ab$ due to two-dimensional superconducting nature. To estimate an upper critical field $\mu_0 H_{c2}$, $T_c^{onset}$ was determined using 90 % of resistivity above $T_c$. The estimated $T_c^{onset}$ and applied field are plotted in Fig. 5(c). Using the WHH theory, which gives $\mu_0 H_{c2}(0) = -0.69 T_c (d\mu_0 H_{c2}/dT)|_{T_c}$ [11], the upper critical fields $\mu_0 H_{c2}^{ab}(0)$ and $\mu_0 H_{c2}^{c}(0)$ are estimated to be 55 and 47 T, respectively. Anisotropy at 0 K, $H_{c2}^{ab}(0)/H_{c2}^{c}(0)$, is estimated to be ~1.17. Small anisotropy is consistent with the previous reports on transport properties of $FeTe_{1-x}Se_x$ and $FeTe_{1-x}S_x$ [12,13].

To discuss the change in crystal structure, lattice constant was determined using x-ray diffraction. The obtained x-ray profile showed only (00$l$) reflections as observed in the as-grown crystals [4]. Lattice constant $c$ for both as-grown and oxygen-annealed $FeTe_{1-x}S_x$ crystal is plotted in fig. 6 as a function of $x$. The lattice constant linearly decreases with increasing S concentration. For all oxygen-annealed crystals, the shrinkage of lattice is observed, as observed in the polycrystalline samples [7]. Although the lattice constant is compressed for $x = 0.03$ and 0.048 as well, the magnetic transition temperature does not change by oxygen annealing. The binding of oxygen in $FeTe_{1-x}S_x$ would be weak, because superconductivity in $FeTe_{1-x}S_x$ induced by oxygen annealing is reversible; superconductivity is suppressed when the superconducting $FeTe_{1-x}S_x$ is annealed in vacuum. Then, reannealing in oxygen reproduces superconductivity in $FeTe_{1-x}S_x$. On the basis of these facts, we think that the oxygen is not substituted for the Te site and would be intercalated into the interlayer site. The evolution of superconductivity by oxygen annealing might be related to a suppression of magnetic moment of excess Fe, which exists at interlayer site and destroys superconductivity [14-17]. To clarify the intrinsic role of oxygen in $FeTe_{1-x}S_x$, studies sensitive to local structure are required.

Figure 7 is a phase diagram of oxygen-annealed $FeTe_{1-x}S_x$ single crystals grown using the self-flux method. Due to the low solubility limit of S for the Te site, we did not obtain the crystals with $x > 0.12$ using this growth method. The antiferromagnetic phase is suppressed, and superconducting phase appears, with increasing S concentration. However, oxygen annealing is required to induce superconductivity for $FeTe_{1-x}S_x$ in which the log-range magnetic ordering is suppressed. In fact, both two factors are required for the appearance of superconductivity in $FeTe_{1-x}S_x$. One factor is S substitution for Te, which suppresses the long-range antiferromagnetic ordering of Fe-square lattice. Another factor is oxygen, which would suppress the local moment of excess Fe that interferes with superconductivity.

IV. Conclusion

We synthesized FeTe$_{1-x}$S$_x$ single crystals using the self-flux method. By oxygen annealing, lattice constant *c* for all crystals was decreased. Oxygen annealing induced superconductivity in FeTe$_{1-x}$S$_x$ crystals, only when the antiferromagnetic transition had been suppressed by S substitution. Anisotropy of superconductivity in oxygen-annealed FeTe$_{0.886}$S$_{0.114}$ was estimated to be 1.17. The phase diagram of oxygen-annealed FeTe$_{1-x}$S$_x$ single crystal was established. To realize superconductivity in FeTe$_{1-x}$S$_x$, both S substitution enough to suppress the antiferromagnetism and oxygen annealing are required.

Acknowledgement

This work was partly supported by Grant-in-Aid for Scientific Research (KAKENHI).

Figure captions

Fig. 1. Optical-microscope image of oxygen-annealed FeTe$_{0.886}$S$_{0.114}$ single crystals.

Fig. 2. (a)Temperature dependence of resistivity for Fe$_{1.08}$Te and as-grown FeTe$_{1-x}$S$_x$. (b) Temperature dependence of resistivity for oxygen-annealed FeTe$_{1-x}$S$_x$.

Fig. 3. Enlargement of the resistive superconducting transitions for oxygen-annealed FeTe$_{1-x}$S$_x$.

Fig. 4. Temperature dependence of magnetic susceptibility for oxygen-annealed FeTe$_{1-x}$S$_x$.

Fig. 5 (a) Temperature dependence of in-plane resistivity for FeTe$_{0.886}$S$_{0.114}$ under magnetic fields up to 7 T perpendicular to $c$ axis. (b) Temperature dependence of in-plane resistivity for FeTe$_{0.886}$S$_{0.114}$ under magnetic fields up to 7 T parallel to $c$ axis. (c) Temperature dependence of upper critical field $\mu_0 H_{c2}$.

Fig. 6. S concentration dependence of lattice constant $c$ for both as-grown and oxygen-annealed FeTe$_{1-x}$S$_x$ crystal.

Fig. 7. Phase diagram of oxygen-annealed FeTe$_{1-x}$S$_x$ single crystal grown using the self-flux method. Antiferromagnetic, paramagnetic and superconducting are abbreviated as AFM, PM and SC. AFM transition temperature ($T_N$) was determined from resistivity measurement.

Fig. 1.

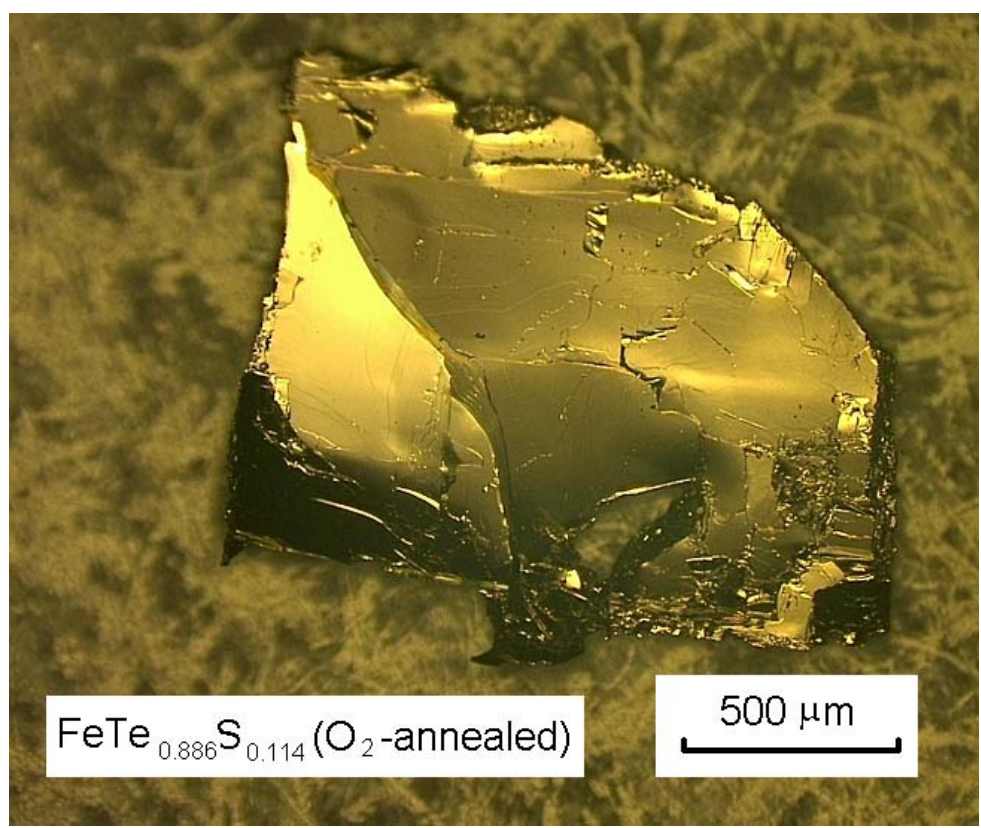

Fig. 2.

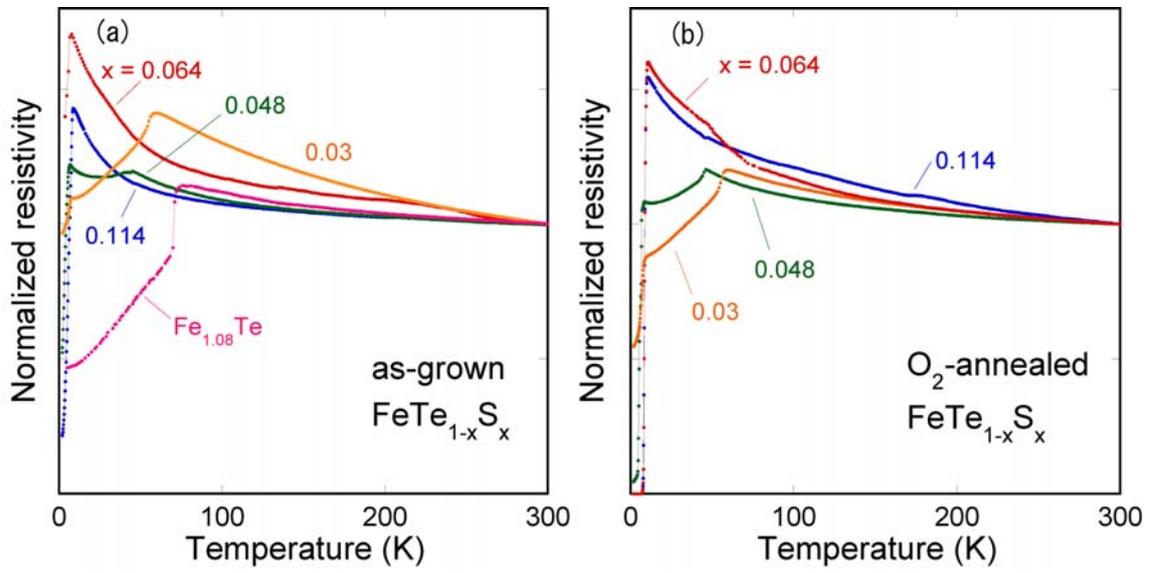

Fig. 3.

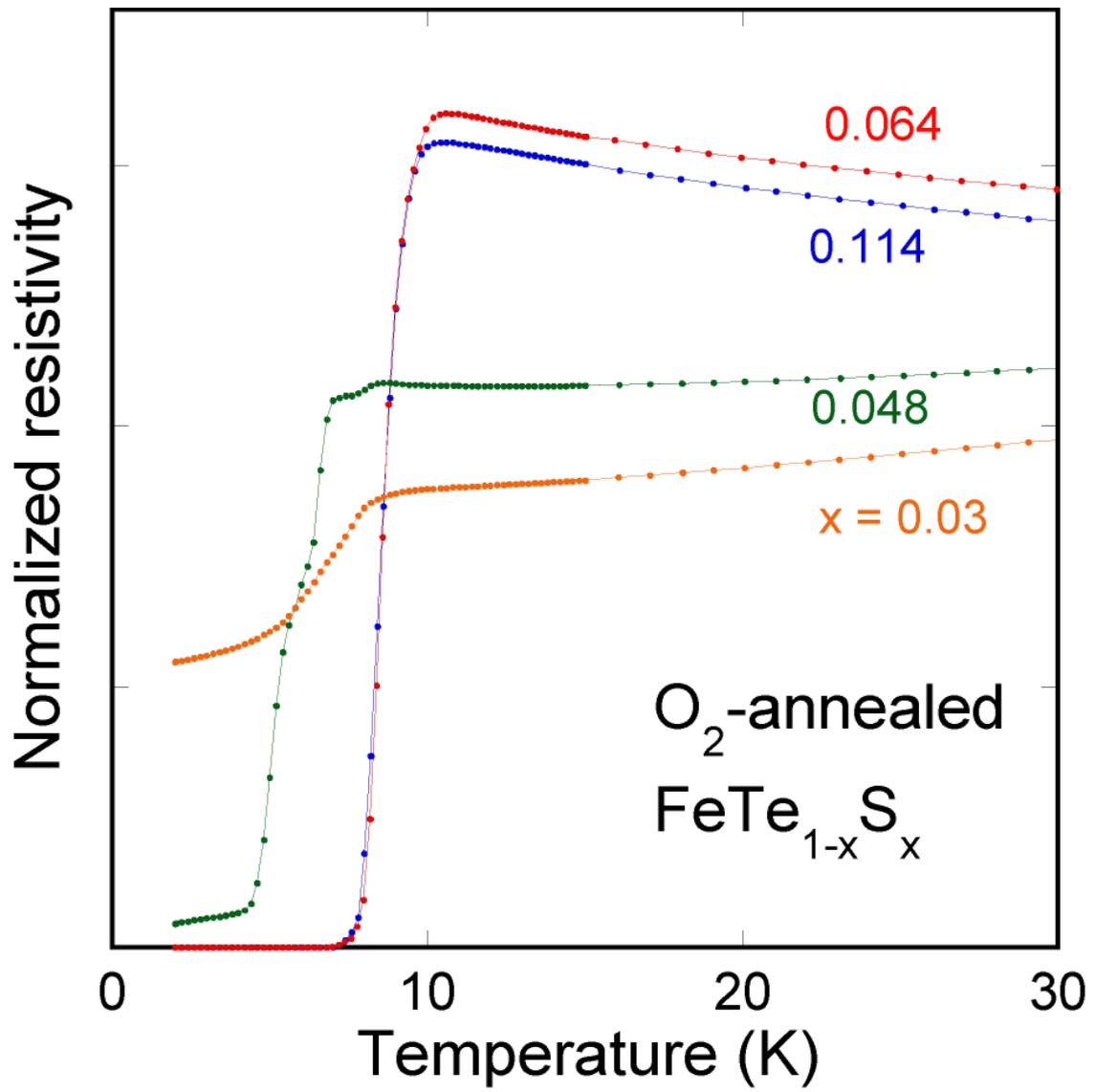

Fig.4.

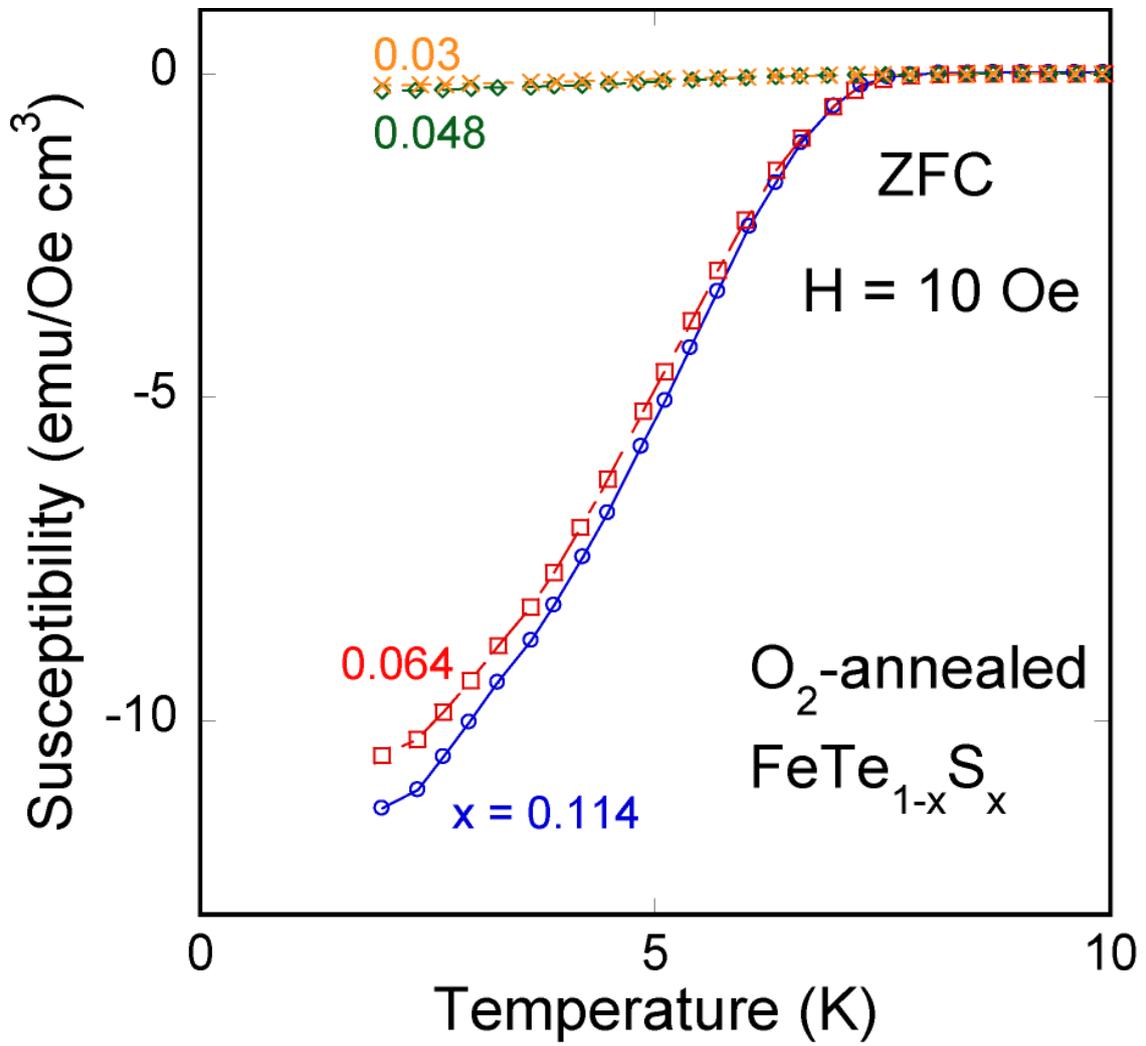

Fig. 5.

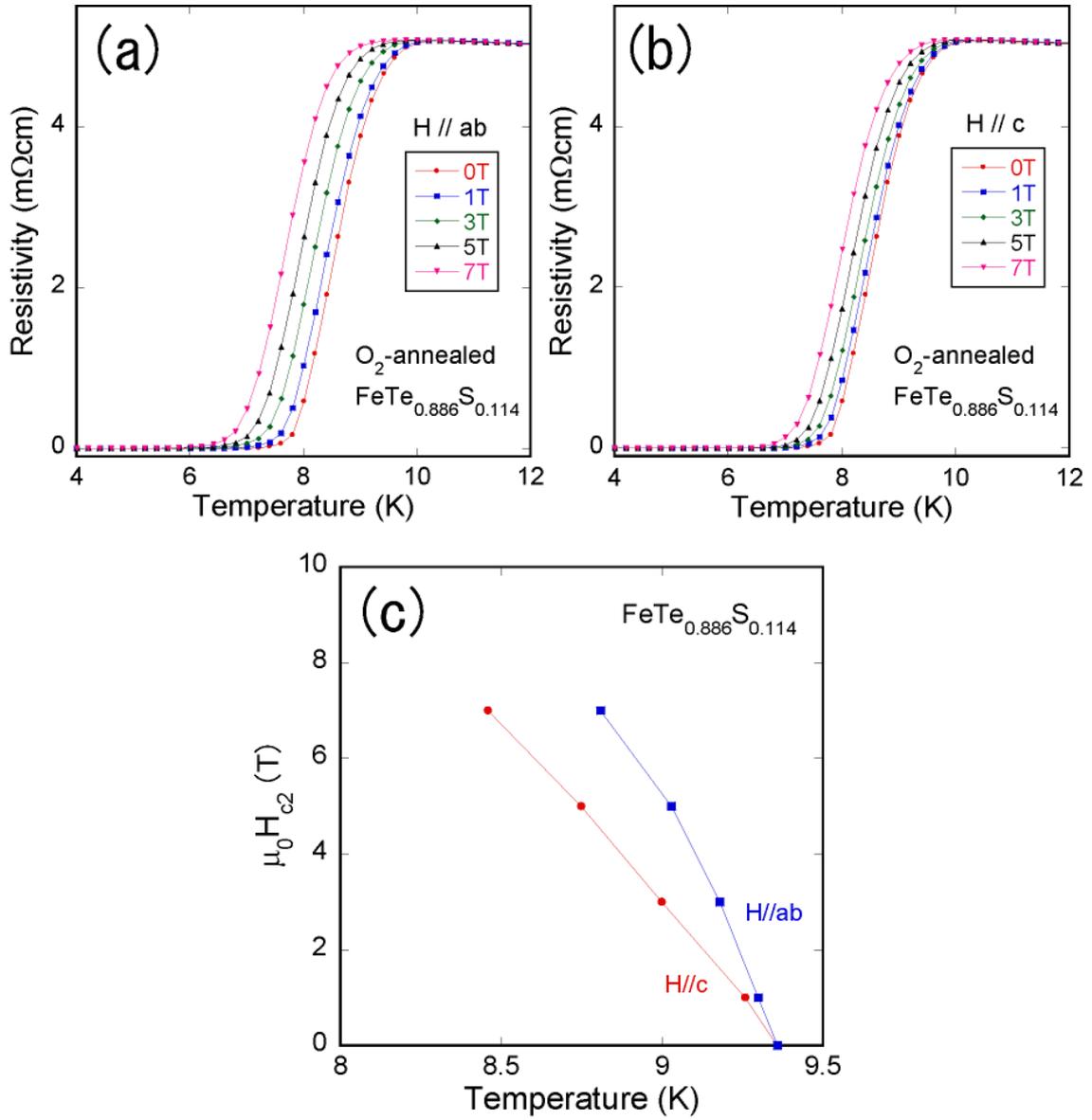

Fig. 6.

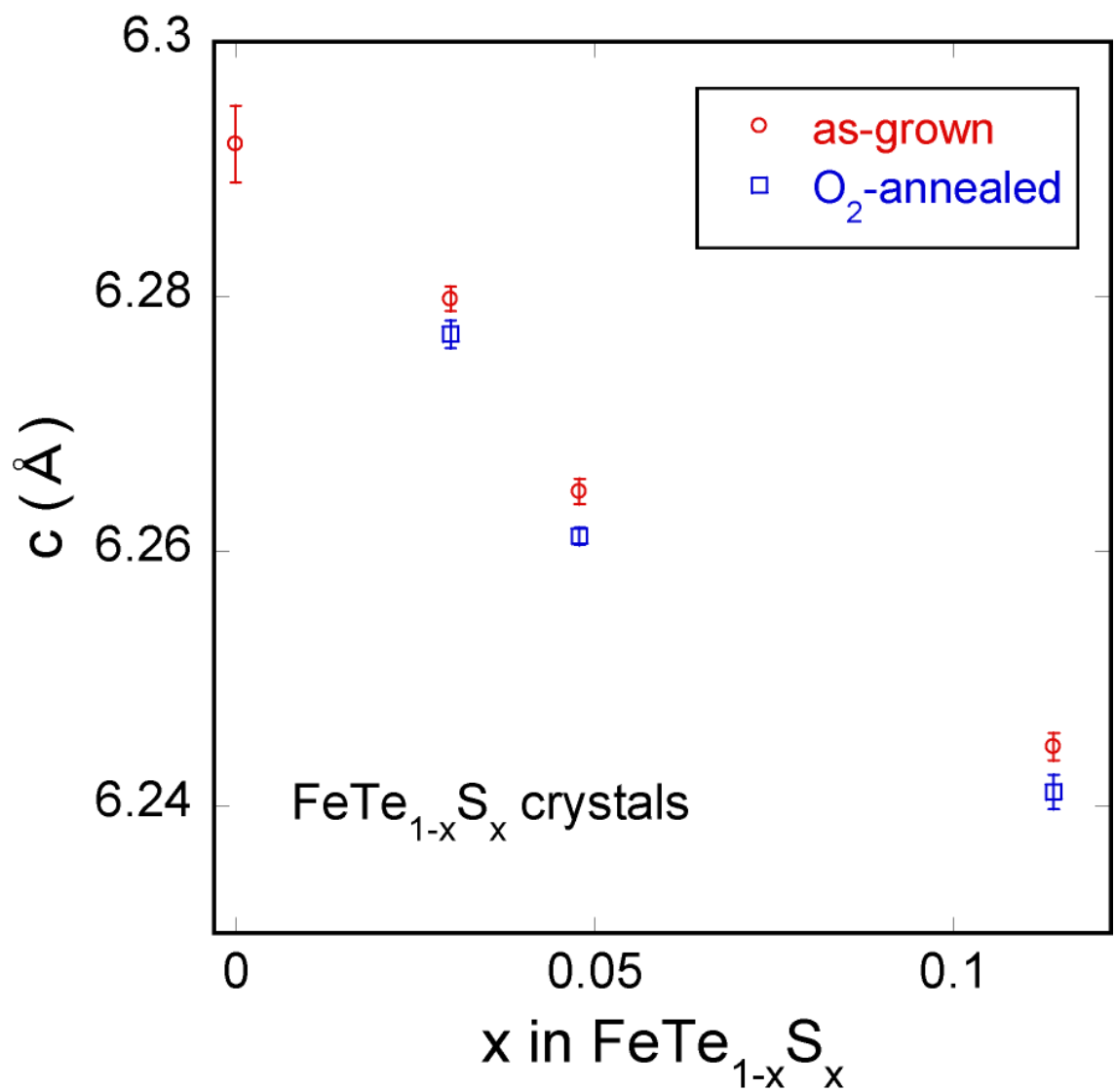

Fig. 7.

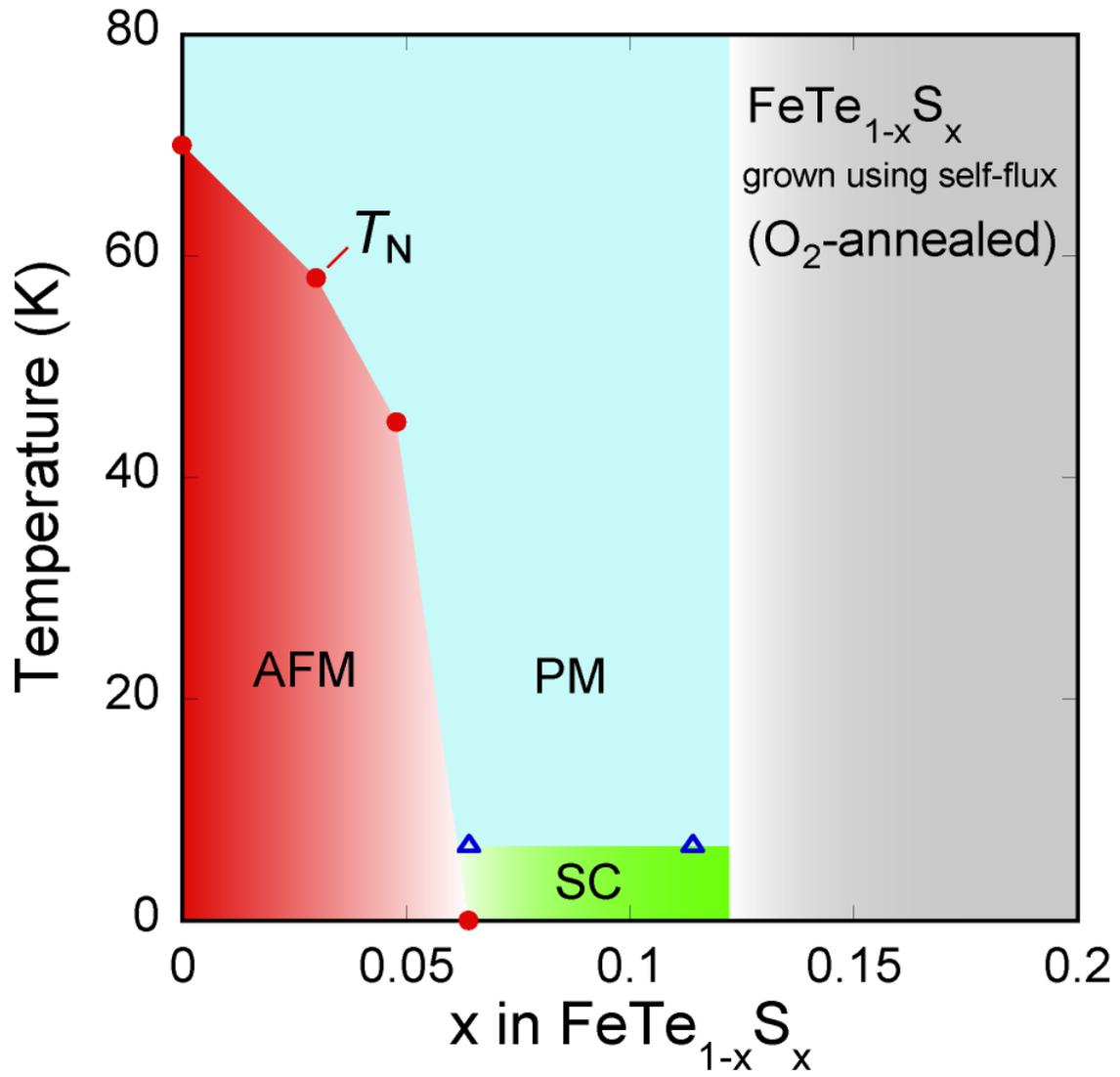